\documentclass[aps,prl,twocolumn,showpacs,superscriptaddress]{revtex4}

\usepackage{graphicx}
\usepackage{amsmath,amssymb}
\usepackage{overpic}
\usepackage{xcolor}

\newcommand*{\abs}[1]{\left|#1\right|}

\newcommand*{\ellp}{\ell_{\rm P}}
\newcommand*{\ellk}{\ell_{\rm K}}

\newcommand{\bstar}{b^\ast}
\newcommand{\cstar}{c^\ast}
\newcommand{\dstar}{d^\ast}

\newcommand{\sgn}{\operatorname{sgn}}

\begin{document}

\title{Confined polymers in the extended de Gennes regime}
\author{E. Werner}

\author{B. Mehlig}
\affiliation{Department of Physics, University of Gothenburg, Sweden}
\date{\today}

\begin{abstract}
We show that the problem of describing the conformations of a 
semiflexible polymer confined to a channel 
can be mapped onto an exactly solvable model in the so-called
extended de Gennes regime. This regime (where the polymer is neither weakly nor
strongly confined) has recently been studied intensively experimentally and by means of computer simulations. The exact solution predicts precisely how the conformational fluctuations depend upon the channel width and upon the microscopic parameters characterising
the physical properties of the polymer. 
\end{abstract}
\pacs
{87.15.A-, 36.20.Ey, 05.40.Fb, 87.14.gk}

\maketitle

The conformations of a polymer change substantially when it is confined to a channel that is significantly smaller than
its unconfined radius of gyration, because it must extend strongly in the channel direction.
How this extension depends on the channel width $D$ and on the physical properties of the polymer
(its Kuhn length $\ellk$, effective width $w$, and contour length $L$) is a question with a very long history. In recent years there has been a surge of interest in this question, stimulated by the possibility of using nanochannels as a tool to study and manipulate DNA molecules \cite{tegenfeldt2004,reisner2005,reisner2007,persson2009,persson2012,werner2012,zhang2013}, see Ref.~\cite{reisner2012} for a review.
Corresponding model systems (worm-like chains with short-range repulsive interactions)
have been intensively studied by computer simulations \cite{wang2011,werner2012,tree2013,dai2013,muralidhar2014,dai2014}.

At present our understanding of this question relies upon computer simulations and mean-field arguments 
\cite{deGennesBook,morrison2005,brochard-wyart2005,odijk2008,wang2011,werner2012,dai2013,tree2013,dai2014}.
For large channel widths and very long polymers the statistical properties of the polymer conformations are
commonly analysed by dividing the confined polymer globule into a series of smaller spherical blobs of size $D$. 
One assumes that Flory's mean-field scaling results holds for each blob (that its extension is proportional
to $L_{\rm blob}^{3/5}$ where $L_{\rm blob}$ is the length of the polymer in the blob),
and concludes that
the extension of the confined polymer globule is proportional to $D^{-2/3}$ \cite{deGennesBook}.
This regime is known as the \lq de Gennes regime\rq{}. 

Closely related mean-field arguments were invoked to describe conformational fluctuations of nano-confined DNA in the
\lq extended de Gennes regime\rq{}. This regime is  characterised by $\ellk \ll D \ll \ellk^2/w$ for a self-avoiding worm-like chain of width $w$.  Based on a Flory-type mean-field argument one estimates that the extension scales as $\propto D^{-2/3}$ \cite{odijk2008,wang2011,dai2013,dai2014}.
But we know that at least for the unconfined polymer Flory's mean field theory is not exact.
That the Flory exponent $\nu=3/5$ is close to the exact exponent $0.588$
\cite{zj1977,zj1998,clisby2010} is the result of the cancellation
of two errors, see Ref.~\cite{BG1990} for a review of this well-known fact.

In this Letter we show that the problem of describing the conformations of a confined worm-like chain in the extended de Gennes regime
can be mapped onto a one-dimensional model that has been rigorously analysed by means of large-deviation theory \cite{vanderhofstad2003}.
This is important since a large number of recent experiments and computer simulations
determining the conformational fluctuations of long confined DNA 
pertain to this regime.  

Our mapping makes it possible 
to determine exactly how the conformational fluctuations 
depend upon the channel width and upon the microscopic parameters 
characterising the physical properties of the polymer. 
The mapping translates the asymptotic results of Ref.~\cite{vanderhofstad2003} into a precise picture for the conformations of long polymers in the extended de Gennes regime, obtained as an \lq optimal path\rq{} in the limit of large values of $L$. 
This permits us, first, to compute the distributions of the end-to-end distance of the polymer and of its extension.
The distributions are non-Gaussian and asymmetric, in excellent agreement with Monte-Carlo simulations of
   a self-avoiding confined polymer. Computation of the low-order moments shows 
that mean-field approximations~\cite{odijk2008,wang2011,dai2014} give the {\em exact}  scaling for the extension
(unlike Flory theory for an unconfined polymer globule). Second, 
the one-dimensional random-walk model exhibits a universal scaling relation.
%%  that allows to map results for microscopically different physical systems 
%% at different parameter values onto a universal model in the extended de Gennes regime.
 For the polymer in the extended de Gennes regime this implies 
 that any results derived for one set of parameters (e.g. by simulations) 
 can be directly translated to any other set of parameters within the extended de Gennes regime.
Third, the mapping makes it possible to compute how the conformational fluctuations in the bulk differ from
those at the right or left end of a polymer globule in a channel. Fourth, the theory shows that self-avoidance is a singular perturbation: the limit $w\rightarrow 0$ does not
correspond to ideal fluctuations. Fifth, the system exhibits a phase transition: there is a critical end-to-end distance below
which the conformations of the polymer change qualitatively.

{\em Mapping to random-walk model.} We model the polymer as a freely jointed chain of $N$ 
units with excluded volume $v$ and Kuhn length $\ellk$, confined to a channel (cross section $D^2$) 
that extends in the $z$-direction. For $D \gg \ellk$  the macroscopic fluctuations of a dilute polymer depend 
on the microscopic details of the model only through $N$, $v$, and $\ellk$ \cite{grosberg1994}.
In this case it suffices to analyse a freely jointed chain of spherical beads 
of diameter $a$, excluded volume $v=4\pi a^3/3$, connected by ideal rods of length $\ellk$. 
We now show that the statistics of the $z$-components can be mapped to a random-walk model.
The $z$-coordinates of an ideal polymer ($a=0$, $v=0$) obey the statistics of a simple random walk with independent steps of mean zero and variance $\sigma_0^2=\ellk^2/3$. To obtain a statistically correct ensemble $P(\{z_n\})$ for the $z$-coordinates of a self-avoiding polymer, one generates an ensemble $P_{\rm ideal}(\{z_n\})$ of ideal polymers, and removes all configurations with overlaps:
\begin{align}
\label{eqn:PFactorisation_z}
P(\{z_n\}) \propto P_{\rm ideal}(\{z_n\}) A(\{z_n\})\,.
\end{align}
Here the acceptance function $A(\{z_n\})$ is the fraction of ideal configurations that are free of overlaps. 
It is in general a complicated object, yet if the excluded volume $v$ is small enough, it becomes quite simple. In this limit, most collisions occur between monomers with a contour separation $\abs{n-m}$ that is so large that the $x$- and $y$-coordinates of the two monomers are independent of each other. Their collision probability is simply proportional to the probability that their $z$-coordinates are close to each other.
Since the collision probability only depends on the return probability of the $z$-coordinate, the problem is effectively one-dimensional. Now imagine dividing the $z$-axis into bins of size $\epsilon$, where $a\ll \epsilon\ll \ellk$.
For two monomers to collide, their $z$-coordinates must lie in the same bin. 
Supposing that they do, we write the probability for the monomers to collide as $2\eta/\epsilon$, the constant of proportionality $\eta$ 
is discussed below. Assuming that $\eta/\epsilon\ll 1$ the factor $A(\{z_n\})$ in Eq.~(\ref{eqn:PFactorisation_z}) can be written as
\begin{align}
\label{eqn:A_factorises}
A(\{z_n\}) =\!\!\!\!\!\!\!\!\! \prod_{
	\substack{1\le n< m\le N\\
				z_n=z_m}
} \!\!\!\!\!\!\!\!
(1\!-\!2\eta/\epsilon)&\!=\!
 \exp\bigg[\!\!-\frac{2\eta}{\epsilon}\!\!\!\sum_{1 \le n<m \le N}\!\!\!\!\!\! \!\!\delta^{(\epsilon)}_{z_n,z_m}\bigg]\,.
\end{align}
Here ${\delta}^{(\epsilon)}_{z_n,z_m}$ is unity if $z_n$ and $z_m$ fall into the same bin,
%  of width $\epsilon$, 
and zero otherwise.

The problem of describing the conformations of the confined polymer is thus mapped to that of a random walk on $\mathbb{Z}$, consisting of $N$ steps 
with variance $\sigma_0^2/\epsilon^2$. Each time two steps land on the same integer, the  walk incurs a \lq penalty\rq{} of $2\eta/\epsilon$.
This problem is known as the weakly self-avoiding random walk or the Domb-Joyce model in one dimension \cite{vanderhofstad2003}. 
It remains to establish the range of validity of our assumptions. Provided that $D\gg \ellk$, 
the $x$- and $y$-coordinates of two monomers of an ideal polymer are all statistically independent of each other if their contour separation $\abs{n-m}$ exceeds $D^2/\ellk^2$ \cite{werner2013}. 
For our mapping to be valid most collisions must occur between monomers beyond this separation. This requires that an ideal blob of $k=D^2/\ellk^2$ monomers is free of collisions. 
We estimate that this is the case provided $k^2v/(k\ellk^2)^{3/2}\ll 1$, in other words $D v /\ellk^4\ll 1$. We now show that these conditions correspond precisely
to the extended de Gennes regime for a self-avoiding worm-like chain of width $w$ and persistence length $\ellp$. Such a polymer has a Kuhn length of $\ellk=2\ellp$ and an excluded volume that can be estimated by that of a cylinder \cite{onsager1949}: $v=(\pi/2) \ellk^2 w$ (assuming $w\ll\ellk$). 
In summary our mapping is valid in the extended de Gennes regime $\ellk\ll D\ll \ellk^2/w$.

Finally we derive an explicit expression for $\eta$ in Eq.~(\ref{eqn:A_factorises}). Consider two monomers with $z$-coordinates
in the same bin. For a long polymer the monomers are typically not close to either of the ends of the polymer.
In this case and provided that $D\gg \ellk$, the probability that the monomers collide is determined by the 
equilibrium density \cite{werner2013} of the $x$- and $y$-coordinates $\rho(x,y)=(2/D)^2 \sin^2(\pi x/D)\sin^2(\pi y/D)$
in a square channel of width $D$:
\begin{equation}
\label{eqn:etaExpression}
\eta
=v/2\, \int_0^D\!\!{\rm d}x\,{\rm d}y\,\rho^2(x,y)
= \frac{9}{8} \frac{v}{D^2}\,.
\end{equation}
This equation shows how the parameter $\eta$ depends on
the channel width $D$ and, through the excluded volume $v$,
on the microscopic parameters of the polymer.
\begin{figure}
\begin{overpic}[width=\columnwidth]{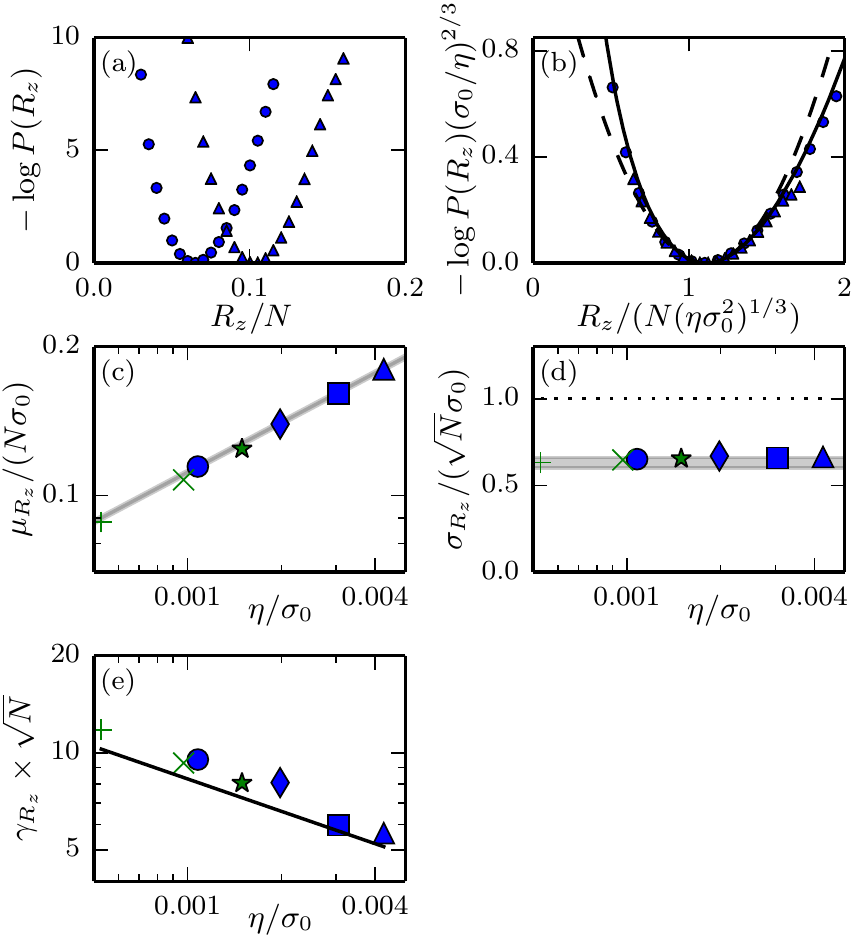}
\put(58,5){\includegraphics[width=3cm,clip]{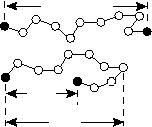}}
\put(66, 29){\colorbox{white}{$R_z=r_z$}}
\put(68, 5){\colorbox{white}{$R_z$}}
\put(65, 11){\colorbox{white}{$r_z$}}
\put(85, 8){\colorbox{white}{\small (f)}}
\end{overpic}
% \includegraphics[width=\columnwidth]{fig1.pdf}
% \raisebox{13cm}{\begin{overpic}[width=3cm,clip]{fig2f.jpg}
% \put(25, 77){\colorbox{white}{$R_z=r_z$}}
% \put(31, 0){\colorbox{white}{$R_z$}}
% \put(21, 19){\colorbox{white}{$r_z$}}
% \end{overpic}}
\caption{{(color online)}.
{\bf a} Distribution $P(R_z)$ of the extension $R_z$ for 
$D/\ellk = 4, a/\ellk = 0.14$ ($\circ$), $a/\ellk = 0.22$ ($\triangle$).
{\bf b} Same distributions plotted as a function of
the scaled variable $R_z/[N(\eta\sigma_0^2)^{1/3}]$, Eqs.~(\ref{eq:PR}) and (\ref{eq:SR}).
Also shown is the asymptotic exact solution $J(b)$
(solid line, see text) as well as a Gaussian approximation (dashed line). 
{\bf c} Scaling prediction  for the mean extension
[Eq.~(\ref{eqn:extensionGeneral})], shaded grey area. Its width
corresponds to the uncertainty of the coefficient $b^\ast$ given in the text below
Eq.~(\ref{eqn:varianceGeneral}). Also shown are results of
Monte-Carlo simulations (symbols) for 
$D/\ellk = 4$, $a/\ellk = 0.14$ ($\circ$), $0.17$ ($\Diamond$), $0.20$ ($\Box$), $0.22$ ($\triangle$) and $D/\ellk = 6$, $a/\ellk = 0.14$ ($+$),
$0.17$ ($\times$), $0.20$ ($\ast$).
{\bf d} Standard deviation of the extension distribution from simulations (symbols), compared to Eq.~(\ref{eqn:varianceGeneral}), shaded grey area. The dotted line indicates the variance of the corresponding ideal polymer.  {\bf e} Skewness of the extension distribution from simulations (symbols) and theory (\ref{eqn:skewnessGeneral}, solid line).
{\bf f} Illustrates a conformation
that satisfies the bridge condition (top) and a conformation that does not (bottom). See text.}
\label{fig:extensionAndVariance}
\end{figure}
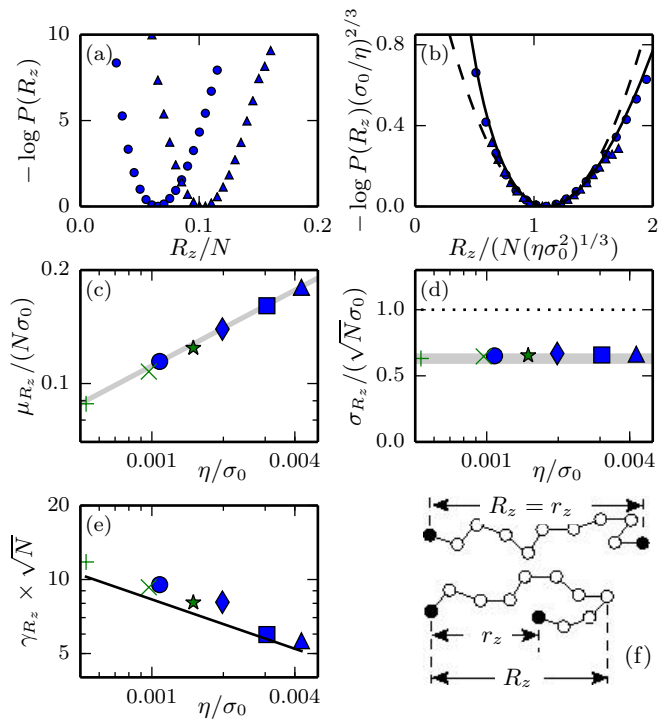

{\em Exact results for the extended de Gennes regime.}
The mapping to a one-dimensional random-walk problem is the main result of this Letter. It is significant because exact asymptotic solutions
have been obtained for this model in the limit of small $\eta$.  These results allow us to rigorously analyse the statistical properties of a self-avoiding polymer confined to a channel,
provided that $\eta/\sigma_0 \ll 1$.
Eq.~(\ref{eqn:etaExpression})  shows that this condition is met in the extended de Gennes regime.
In this case a scaling relation holds for the one-dimensional 
weakly self-avoiding random walk \cite{vanderhofstad2003},
derived from the invariance of $P_{\rm ideal}$ under $z \rightarrow \alpha z$ and $n \rightarrow 
\alpha^2 n$, and the form (\ref{eqn:A_factorises}) of $A$.
For the confined polymer, this relation implies that plotting results in terms of the scaled variables
\begin{equation}
 \label{eqn:scaling}
n' = n\,  (\eta/\sigma_0)^{2/3} \quad \mbox{and} \quad z' = (z/\sigma_0) \, (\eta/\sigma_0)^{1/3}
 \end{equation}
must give rise to universal laws. 
But much more can be deduced. In the limit of large $N$ the random-walk model admits an exact asymptotic solution
that can be obtained by large-deviation theory \cite{vanderhofstad1998,vanderhofstad2003,denhollander2009}. 
This fact has important implications for the physics of confined polymers. 
Many remain to be worked out, but
in the following we show what the exact solution tells us about
a number of important questions concerning confined polymers in the extended de Gennes regime.

{\em Extension.} The distribution of the extension $R_z = \max_i{z_i}-\min_i{z_i}$ of the polymer in the channel
acquires a large-deviation form in the limit of large $N$
\begin{equation}
\label{eq:PR}
P(R_z) \sim \exp[ -N S(R_z/(N\sigma_0),\eta/\sigma_0)]\,.
\end{equation}
 Eq.~(\ref{eqn:scaling}) implies that the \lq action\rq{} $S$ obeys the scaling law 
\begin{equation}
\label{eq:SR}
S\Big(\frac{R_z}{N\sigma_0},\frac{\eta}{\sigma_0}\Big) =
\Big (\frac{\eta}{\sigma_0}\Big)^{2/3}\, J\Big[\Big(\frac{\eta}{\sigma_0}\Big)^{-1/3} \,\frac{R_z}{N\sigma_0}\Big]\,,
\end{equation}
where $J(b)$ is a scaling function.
To compare the scaling predictions (\ref{eq:PR}) and (\ref{eq:SR}) to numerical results
we have performed Monte Carlo simulations for a freely jointed chain of spherical monomers of diameter $a$, with excluded volume $v = 4 \pi a^3/3$.
To determine its conformations when confined to a channel we
used the Metropolis algorithm with crankshaft trial updates \cite{yoon1995} and performed simulations for $D=4\ellk$ and $D=6\ellk$.
The inequality $\ellk \ll D$ is only weakly satisfied.  In order to nevertheless obtain a good estimate for $\eta$
we used a numerical solution for the equilibrium density $\rho(x,y)$ of the ideal freely jointed chain.
We measured the distribution of $R_z$. It is shown in panel {\bf a} of Fig.~\ref{fig:extensionAndVariance} as a function of $R_z$ for two different sets of parameters.
Panel {\bf b} shows that the data collapse upon scaling $x$- and $y$-axis according
to (\ref{eq:SR}). We computed the universal scaling function, $J(b)$, by numerical solution of an eigenvalue problem
(Eq. (3.49) in Ref.~\cite{denhollander2009} and Eq. (0.15) in Ref.~\cite{vanderhofstad1995}). 
The result is shown 
in Fig.~\ref{fig:extensionAndVariance}{\bf b} as a solid line. We observe excellent agreement far into the 
non-Gaussian tails of the distribution.

The function $J(b)$ determines the statistical properties of the extension.
The location $b^\ast$ of the minimum of $J$ gives the mean $\mu_{R_z}$ of $R_z$, 
and the curvature $1/{c^\ast}^2$  at
this point yields the corresponding variance $\sigma_{R_z}^2$:
\begin{align}
\mu_{R_z}/N &= \bstar \sigma_0^{2/3}\eta^{1/3}\,, \label{eqn:extensionGeneral} \\
\sigma_{R_z}/\sqrt{N} &= \cstar \sigma_0\,. \label{eqn:varianceGeneral}
\end{align}
The third derivative
of $J$ at $b^\ast$ determines the sknewness $\gamma_{R_z}$ of the distribution:
\begin{align}
\gamma_{R_z} \sqrt{N}&= \dstar ( \eta/ \sigma_0)^{-1/3}\label{eqn:skewnessGeneral}\,,
\end{align}
showing that $P(R_z)$ is not Gaussian. Rigorous bounds for the universal parameters $\bstar$ and $\cstar$ 
were obtained in Refs.~\cite{vanderhofstad1998,vanderhofstad2003}: $1.104 \!\leq \!\bstar \!\leq\! 1.124$, and  $ 0.60 \!\leq \!\cstar \!\leq\! 0.66$.
From our numerical solution for $J(b)$ we find $\dstar = 0.83$. 
For a self-avoiding worm-like chain Eqs.~(\ref{eqn:extensionGeneral}) to (\ref{eqn:skewnessGeneral}) yield:
\begin{align}
\mu_{R_z}/L &=0.9338(84)\big(\ellk w/ D^2\big)^{1/3} \label{eqn:extensionWLC}\,,\\
{\sigma_{R_z}}/{(L\ellk)^{1/2}} &= 0.364(17)\,, \label{eqn:varianceWLC}\\
{\gamma_{R_z}} {(L/\ellk)^{1/2}} &=0.57 \big(\ellk w/ D^2\big)^{-1/3}  \label{eqn:skewnessWLC}
\end{align}
Eq.~(\ref{eqn:extensionWLC}) implies that the Flory mean-field arguments of Refs.~\cite{odijk2008,wang2011,dai2013}
give the {\em exact}  scaling exponent in the extended de Gennes regime. For an uncofined polymer, by contrast, Flory's mean-field theory does not give the exact exponent.
Recent simulations by Dai {\em et al.} \cite{dai2014} measure the extension of a 
freely jointed chain with cylindrical monomers
in a channel. Their results are consistent with the power laws in Eqs.~(\ref{eqn:extensionWLC}) and (\ref{eqn:varianceWLC}), 
but with numerical prefactors $0.84$ and $0.37$, respectively. The discrepancy for the extension is probably due to the
fact that $\ellk \ll D$ holds only approximately for some of the channels used,
resulting in a correction to Eq.~(\ref{eqn:etaExpression}).

{\em End-to-end distance}. The distribution $P(r_z)$ of the end-to-end distance $r_z=|z_1-z_N|$ is in general
different from the distribution of the extension, and this difference reveals interesting physics. $P(r_z)$
is of the form (\ref{eq:PR},\ref{eq:SR}) save for a different scaling function $I(b)$ that replaces $J(b)$.
Let is discuss similarities and differences between these scaling functions. First, 
one finds that the locations of the minima of $I$ and $J$ and the curvatures at this point
are identical. This reflects the fact that the mean and variance of the end-to-end distance and the extension must
agree for very long chains that only sample the Gaussian part of the distribution. \
Second, it follows from large-deviation theory that the right tails of the distributions of the extension and the end-to-end distance
are dominated by conformations that fulfil the so-called  \lq{}bridge condition\rq{} 
$r_z = R_z$. An example is given in Fig.~\ref{fig:extensionAndVariance}{\bf f}.
Third, 
the small-$b$ behaviour of $I$ and $J$ differ substantially.
$I(b)$ is linear in $b$ when $b$ is smaller than a critical value $b^{\ast\ast} = 0.85$ 
\cite{vanderhofstad2003a}. 
This implies that the free energy in the extended de Gennes regime 
is a linear function of the end-to-end distance $r_z$
when $r_z < r_z^{\ast\ast}$.
Imagine compressing an extended conformation (top conformation in Fig.~\ref{fig:extensionAndVariance}{\bf f}) by bringing the ends closer together. 
At first the bridge condition remains
satified as work is performed against the entropic  restoring force. 
But there is a critical value $r_z^{\ast\ast}$ where
it becomes advantageous for the polymer to make a U-turn as shown in Fig.~\ref{fig:extensionAndVariance}{\bf f}, bottom. 
From this point on the restoring force is independent of $r_z$ implying infinite compressibility. 

\begin{figure}
\begin{overpic}[width=\columnwidth,clip]{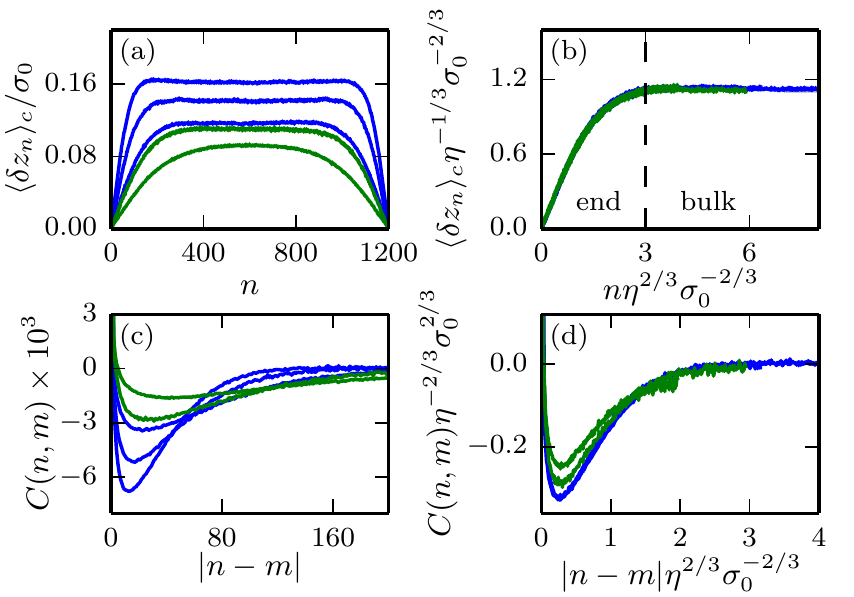}
\end{overpic}
\caption{{(color online)}. {\bf a} Bias (see text) as a function of monomer number. {\bf b} Data in the left half of panel a, 
rescaled to show the universality of the shape. Shaded grey area: prediction from Eq.~(\ref{eqn:extensionGeneral}). The vertical dashed line indicates the boundary between the end and the bulk regions of the polymer globule.
{\bf c} The correlation function $C(n,m)$ (defined in the text) averaged over $n,m$ in the interior of the polymer (the plateau of panel a). 
{\bf d} The rescaled correlation function, showing the universal \lq{}correlation hole\rq{}.
In both panels, blue (green) lines correspond to $D=4\ellk$ ($D=6\ellk$). In order of increasing $\eta$, the monomer diameter is given by 
$a/\ellk=0.14, 0.17$ ({\rm green}); $0.14, 0.17, 0.20$  ({\rm blue}).
\label{fig:biasAndCorrelation}}
\end{figure}

{\em Finite-size effects.} 
% It is expected that 
The conformational
fluctuations within a polymer globule that is confined to a channel  differ in the bulk and 
at either boundaries of the globule. But it is not  known
how large the boundary region is, neither how important these finite-size effects are. 
In the extended de Gennes regime the mapping to the random-walk model 
allows us to answer this question by examining the distribution of the 
steps $\delta z_n= z_{n+1}-z_n$ of the random walk that the $z$-coordinates
of the polymer perform in the channel. The random walk is biased because
the monomers share a tendency to step in the same direction -- for a given realisation.
The bias $\langle \delta z_n\rangle_c\equiv\langle \delta z_n \sgn(z_N-z_1)\rangle$ (where $\sgn(x)$ is the signum function) is
shown in 
Fig.~\ref{fig:biasAndCorrelation}{\bf a}. Fig.~\ref{fig:biasAndCorrelation}{\bf b} demonstrates that all data points collapse to a universal curve when rescaled according to Eq.~(\ref{eqn:scaling}). 
Far from either end the bias must approach the bulk prediction (\ref{eqn:extensionGeneral}). Fig.~\ref{fig:biasAndCorrelation}{\bf b} shows that this is indeed the case. 
Near the end of the polymer the bias is smaller. 
For shorter polymers this end effect may give rise to significant finite-size scaling corrections. 
The bias also implies that the right (left) end of the polymer is most likely close to the
right (left) boundary of the globule in the channel. This fact together with
Eq.~(\ref{eqn:scaling}) allows us to estimate 
the contour length that lies in the boundary region of the globule:
\begin{equation}
\Delta L = 1.4\, (D^4\ellk /w^2)^{1/3}\,.
\end{equation}
The factor $1.4$ is an estimate based on the location of the vertical dashed line in 
Fig.~\ref{fig:biasAndCorrelation}{\bf b}.

{\em Singular limit.} 
As the strength of the self-avoidance tends to zero
the fluctuations of $r_z$ remain distinctly smaller than those of an ideal polymer.
This gives rise to a universal \lq{}correlation hole\rq{}
in the correlation $C(n,m) = (\langle \delta z_n\delta z_m\rangle_c - \langle \delta z_n\rangle_c \langle\delta z_m\rangle_c)/\sigma_0^2$, since $\sigma_{r_z}^2=\sigma_0^2\sum_{n,m=1}^N C(n,m)$.
Fig.~\ref{fig:biasAndCorrelation}{\bf c} shows this correlation function. Fig.~\ref{fig:biasAndCorrelation}{\bf d} shows the rescaled version exhibiting the universal correlation hole.

{\em Outside the extended de Gennes regime.} How does the theory break down as the assumptions $\ellk\ll D \ll \ellk^4/v$ are violated? What happens as $D$ approaches $\ellk$, for a worm-like chain such as DNA? The mapping to a one-dimensional model must still hold, 
yet the calculation of the parameters of the weakly self-avoiding random walk must be modified in two ways. The polymer tends to align with the channel direction and is less prone to switch direction \cite{werner2012}. This increases the step size and variance of the ideal random walk. In addition, the probability of collisions is no longer simply given by the solution to a diffusion equation, but requires a more complicated calculation involving the monomer orientation.
If on the other hand the assumption $D \ll \ellk^4/v$ is violated, then the situation is more complicated. Eq.~(\ref{eqn:PFactorisation_z}) still holds, but collisions between nearby monomers ($\abs{m-n}< D^2/\ellk^2$) have a profound influence on the conformations, so that Eq.~(\ref{eqn:A_factorises}) breaks down.

{\em Polymers confined to slits.} It has been suggested that an extended de Gennes regime may exist also for polymers confined to a slit, provided that the slit height $h$ obeys $\ellk \ll h \ll \ellk^2/w$ \cite{dai2012,taloni2013}. A mapping similar to the one presented here shows that such a regime must exist, described by a weakly self-avoiding walk in two dimensions.  While no rigorous results have as yet been derived for this model, it is certainly easier to analyse than the three-dimensional polymer, and can be simulated much more efficiently.

{\em Acknowledgments}.
Financial support by Vetenskapsr\aa{}det and
by the G\"oran Gustafsson Foundation for Research in Natural Sciences and Medicine are gratefully acknowledged.
The numerical computations were performed using resources provided by C3SE and SNIC.

\end{document}